\documentclass[sn-mathphys]{sn-jnl}% Math and Physical Sciences 
\usepackage{siunitx}
\usepackage{comment}\usepackage{booktabs}\usepackage{amsmath}
\usepackage{amssymb}\usepackage{bm}
\usepackage{mathtools}
\usepackage{siunitx}
\DeclarePairedDelimiterXPP\BigOSI[2]%
  {\mathcal{O}}{(}{)}{}%
  {\SI{#1}{#2}}

\jyear{2021}%

\theoremstyle{thmstyleone}

\theoremstyle{thmstyletwo}\def\d{{\, \rm d}}

\theoremstyle{thmstylethree}%

\begin{document}

\title[Bridging Gaps in the Climate Observation Network]{Bridging Gaps in the Climate Observation Network: A Physics-based Nonlinear Dynamical Interpolation of Lagrangian Ice Floe Measurements via Data-Driven Stochastic Models}

\author[1]{\fnm{Jeffrey} \sur{Covington}}\email{jmcovington@wisc.edu}

\author*[1]{\fnm{Nan} \sur{Chen}}\email{chennan@math.wisc.edu}

\author[2]{\fnm{Monica M.} \sur{Wilhelmus}}\email{mmwilhelmus@brown.edu}

\affil[1]{\orgdiv{Department of Mathematics}, \orgname{University of Wisconsin-Madison}, \orgaddress{\street{480 Lincoln Drive}, \city{Madison}, \postcode{53706}, \state{Wisconsin}, \country{USA}}}

\affil[2]{\orgdiv{Center for Fluid Mechanics, School of Engineering}, \orgname{Brown University}, \orgaddress{\street{184 Hope St}, \city{Providence }, \postcode{02912}, \state{Rhode}, \country{USA}}}

\abstract{Modeling and understanding sea ice dynamics in marginal ice zones relies on acquiring Lagrangian ice floe measurements. However, optical satellite images are susceptible to atmospheric noise, leading to gaps in the retrieved time series of floe positions. This paper presents an efficient and statistically accurate nonlinear dynamical interpolation framework for recovering missing floe observations. It exploits a balanced physics-based and data-driven construction to address the challenges posed by the high-dimensional and nonlinear nature of the coupled atmosphere-ice-ocean system, where effective reduced-order stochastic models, nonlinear data assimilation, and simultaneous parameter estimation are systematically integrated. The new method succeeds in recovering the locations, curvatures, angular displacements, and the associated strong non-Gaussian distributions of the missing floes in the Beaufort Sea. It also accurately estimates floe thickness and recovers the unobserved underlying ocean field with an appropriate uncertainty quantification, advancing our understanding of Arctic climate. }


\maketitle

\section{Introduction}\label{sec1}

Sea ice plays a key role in the Arctic climate system \cite{thomas2017sea, weeks1986growth,weiss2013drift,lepparanta2011drift,maslowski2012future,bhatt2014implications}. It modulates important momentum, heat, and material transfer processes between the ocean and the atmosphere \cite{thomson2018overview,squire2020ocean}. Given the sensitivity of the sea ice cover to global warming trends, the observation and modeling of sea ice are critical for  understanding global climate, including monitoring the drastic changes in the Arctic and assessing possible future climate scenarios.

Earth system models typically characterize sea ice as a continuum with viscous-plastic rheology primarily through ice concentration, volume, and thickness \cite{hibler1979dynamic, tremblay1997modeling, hunke1997elastic}. While this traditional modeling approach yields realistic results at the basin scale, at scales of $\mathcal{O}$(10)km and smaller, sea ice exhibits brittle behavior with the motion of individual fragments deviating from a continuum description. In this case, the discrete element method (DEM) \cite{cundall1979discrete, cundall1988formulation, hart1988formulation}, which characterizes the trajectories of individual ice floes, as opposed to clusters of ice, becomes the natural choice to describe sea ice dynamics.
Compared to continuum models, the modeling of individual floes provides a richer representation of sea ice dynamics through local interactions with the oceanic and atmospheric components \cite{lindsay2004new, tuhkuri2018review}. In addition, since the DEM models are developed under Lagrangian coordinates, there is no need for an advective transport scheme to move floes between grid cells as in continuum models, significantly reducing computational costs. The DEM models can also change the spatial resolution as the geophysical situation requires, allowing greater flexibility in the study of sea ice.

The unique advantages and wide applications of models based on the DEM highlight the need for observational sea ice data within the Lagrangian framework. Observed floe trajectories facilitate the development and calibration of DEM models and provide insight into the evolution of sea ice properties. However, despite the increase in satellite missions and the improved techniques in acquiring remote sensing observations, most existing observational products are based on Eulerian descriptions of the sea ice drift field. Exceptions include the Arctic Ocean Sea Ice Drift Reprocessed \cite{ardhuin2020REPROCESSED} and the Making Earth System data records for Use in Research Environments (MEaSUREs) programs \cite{KwokMEaSUREs}. Yet, these measurements cannot adequately resolve sea ice motion at small scales due to the spatial resolution (31.25km for Arctic Ocean Sea Ice Drift Reprocessed product) or the sampling frequency (3-day interval for MEaSUREs) of the data. On the other hand, in situ field measurements using buoys on ice floe surfaces have provided invaluable information, but trajectories are often sparse and are positioned far from the coastline.

Recently, a new Lagrangian floe tracking algorithm, called the Ice Floe Tracker \cite{lopez2019ice, wolfe1998modis}, was developed and applied to optical satellite images. It creates Lagrangian sea ice measurements in the low-sampled regions between the ice pack and the open ocean, commonly known as the Marginal Ice Zones (MIZ). This data set was the first of its kind in that it provides not only the Lagrangian trajectories but also the floe sizes and geometries together with the angular displacements of floes in the MIZ extending throughout the 21\textsuperscript{st} century. Given the demonstrated link between floe rotation rates and the characteristics of the underlying small-scale ocean eddies in the Beaufort Gyre MIZ, these sea ice floe observations have proved to be essential for recovering the state of the underlying turbulent ocean field, providing a unique insight into the multi-scale nature of the ocean \cite{manucharyan2022spinning}.

\begin{figure}[h]
\centering
\includegraphics[]{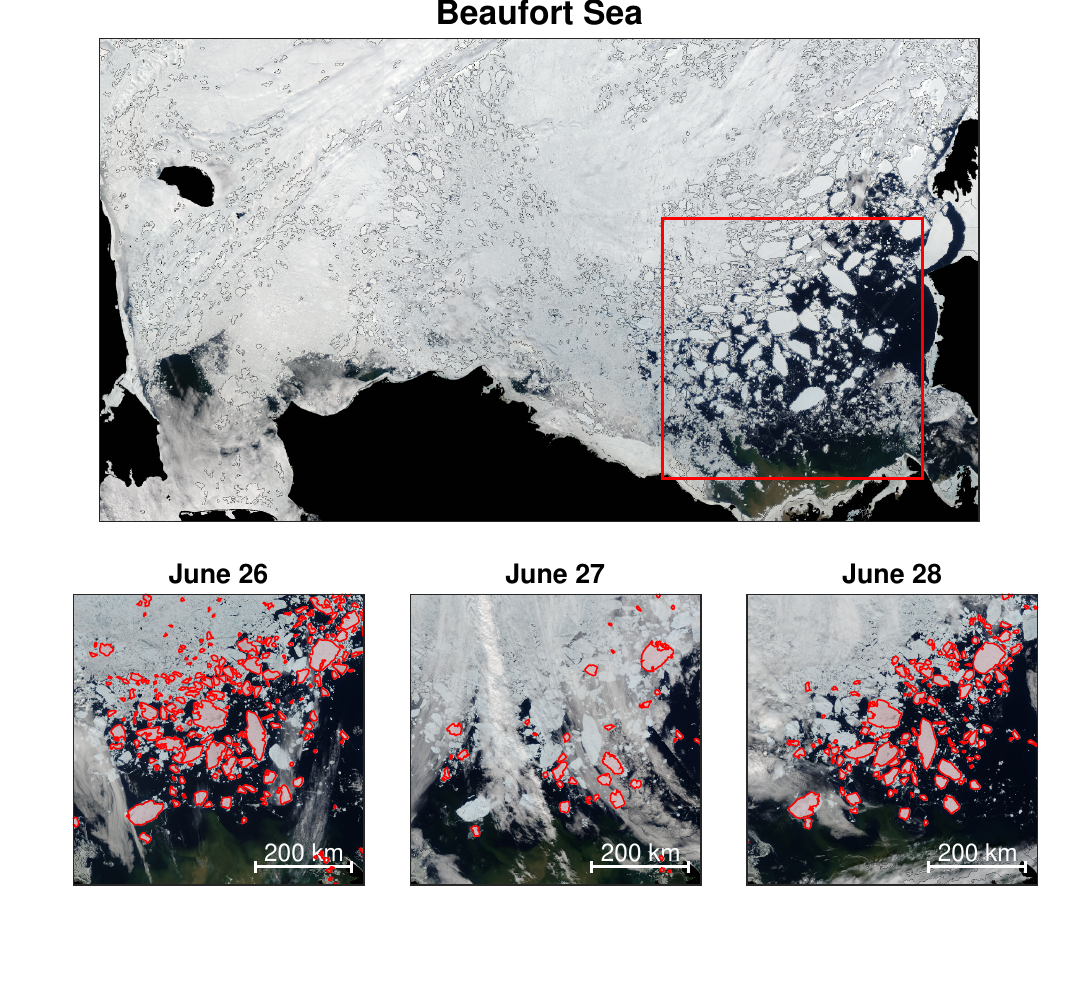}
\caption{Sea ice floes in the Beaufort Sea MIZ. Top panel: Representative Moderate Resolution Imaging Spectroradiometer (MODIS) True Color image (downloaded from the NASA Worldview application) displayed in a WGS 84/NSIDC Sea Ice Polar Stereographic North 70" N projection. For only this figure, the image is oriented 90$^\circ$ from standard Polar Stereographic coordinates so that the top of the image is roughly north. The red box outlines the region of interest. Bottom panels: The observed MIZ of the Beaufort Sea (the box area of the top panel) is shown on three consecutive dates (26.06.2008 to 28.06.2008). Identified floes are contoured with red. On 27.06.2008, the atmospheric noise acts to blur ice floe contours impeding the effective identification of most of the floes. }\label{satellite_fig}
\end{figure}

Atmospheric noise is visible in optical images and leads to many one- or two-day gaps in the retrieved trajectories within the Lagrangian Ice Floe Tracker data set. See Figure \ref{satellite_fig} for an example. A commonly used approach to filling these gaps is to interpolate between the available observations through linear interpolation \cite{manucharyan2022spinning}. Such an approach ignores the MIZ dynamics, leading to trajectories lacking many physical properties. Linear interpolation also fails to retrieve the curvature of the trajectories, which is essential for characterizing the turbulent ocean flow field within the meso/submeso-scale range in polar regions.
On the other hand, physics-based dynamical interpolation incorporates both the available partial observations and knowledge of the coupled atmosphere-ocean-floe dynamics. While the  resulting interpolated trajectories are expected to reflect reality better, traditional dynamical interpolation approaches are often extremely slow and computationally expensive due to the high dimensionality and nonlinearity of the underlying system.

This paper presents an efficient and statistically accurate nonlinear dynamical interpolation framework for recovering the missing floe observations in Lagrangian trajectories. It exploits a balanced physics-based and data-driven construction to address the challenges posed by the high-dimensional and nonlinear nature of the coupled system.  This new method involves a sequential prediction-correction procedure. The error from predicting the missing values in the coupled atmosphere-ocean-floe system is mitigated by incorporating the available observations of floe positions and orientations via Bayesian inference.
One crucial feature of the presented framework is that it exploits a data-driven reduced-order stochastic modeling strategy to advance the statistical forecast of the atmosphere and ocean fields, which are the underlying driving forces of the sea ice motion, but are not considered by the direct curve-fitting algorithms. Particularly, these simple stochastic models describe the time evolution of the leading spectral modes of the atmosphere and ocean fields, where effective stochastic forcing is adopted to characterize the fluctuations at the unresolved scales. Therefore, the resulting surrogate models significantly reduce the computational cost at the forecast step, which is the most time-consuming part in traditional dynamical interpolation approaches.
It is worth highlighting that closed analytic formulae are available for expressing the statistics associated with these simple stochastic models, facilitating the systematic and efficient data-driven model calibration. The calibrated models succeed in accurately predicting the atmosphere and ocean states and the associated uncertainty. The latter is crucial in reaching the least biased state estimate using nonlinear dynamical interpolation, especially in the presence of strong turbulence, which is again completely missed by deterministic curve fitting methods. In addition, the framework allows for simultaneous estimation of several critical physical parameters that cannot be directly inferred from satellite images but are essential for the dynamical interpolation, such as the thickness of each floe, using only relatively short floe trajectories.

The rest of the paper is organized as follows. It starts with the development of the physics-based data-driven dynamical interpolation framework. Then the new method is tested on both a synthetic data experiment and the real data set of floe trajectories in the Beaufort Sea MIZ. The focus here is on the non-interacting floes, but the framework can be easily extended to the interacting ones. The study also includes analysis of the resulting interpolated Lagrangian ice floe trajectories and angular displacements as well as the recovery of several key physical properties of the floes and their associated statistics. The recovered floe trajectory utilizing the traditional linear interpolation approach will serve as a benchmark solution.

\section{Results}\label{sec2}

\subsection{The reduced-order modeling and nonlinear dynamical interpolation framework}
This section presents an overview of the new modeling and nonlinear dynamical interpolation framework consisting of four key steps. The technical details are included in the Methods section and the Supporting Information (SI).

\begin{figure}
\centering
\includegraphics[width=95mm]{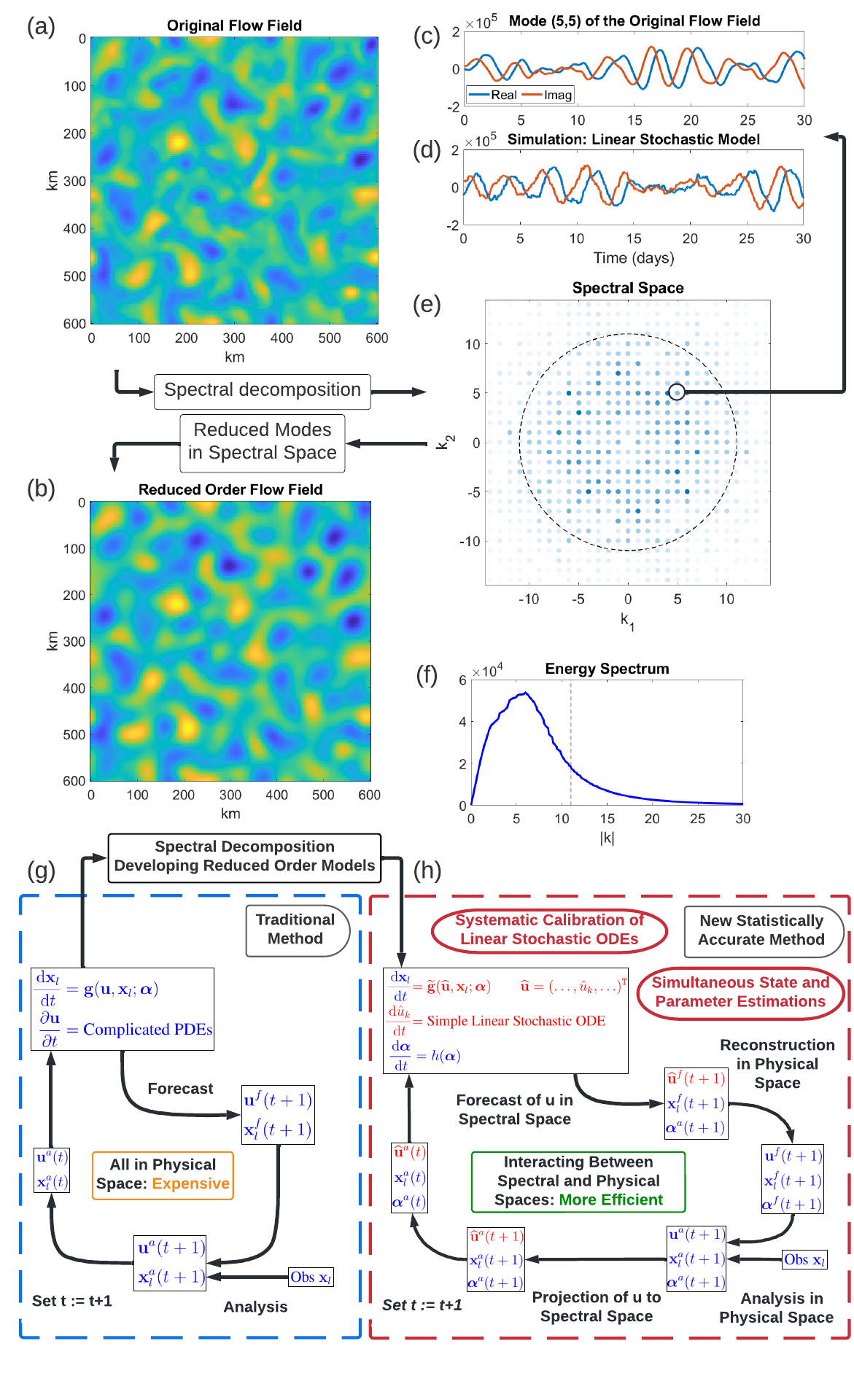}
\caption{Schematic diagram of the new method. Panels (a)--(f):  A spectral decomposition is applied to the output of a complicated ocean model. Only a small set of the most energetic spectral modes are retained. The governing equations of these energetic modes are modeled by the low-cost linear stochastic models, thereby significantly reducing the computational cost. The illustration also compares the true ocean flow field and its reconstructed state. The original field is generated from the two-layer quasi-geostrophic (QG) model, while the reconstructed one only uses modes for which $\vert\mathbf{k}\vert\leq 11$. The top right corner compares the time series of the mode $\mathbf{k}=(5,5)$ associated with the QG model and a random realization from the calibrated linear stochastic model. Panels (g)--(h): The dynamical interpolation is performed via nonlinear data assimilation. The traditional method requires running the original system in the physical space and is extremely expensive. Here, $\mathbf{x}$, $\mathbf{u}$ and $\boldsymbol\alpha$ denote the ice floes, the ocean and the atmospheric state variables, and the model parameters, respectively. In contrast, the new and efficient method for dynamical interpolation alternates between physical and spectral spaces using the reduced-order stochastic models. It has the main benefit of allowing for the simultaneous estimation of state variables and key physical parameters.}\label{schematic_fig}
\end{figure}

This framework works with a coupled atmosphere-ice-ocean system. While this system can take the form of a coupled dynamical model, the framework also allows the atmospheric and/or oceanic components to be given as numerical data. Here, a DEM model is used to characterize the sea ice dynamics, where the individual floe shapes and sizes are drawn from a library of floe observations in the Beaufort Gyre MIZ \cite{lopez2021library}. Given that the observations contain only nearly non-interacting and shape-preserving floes, the ice floe motion can be assumed to be mainly driven by oceanic and atmospheric forcing, which are calculated from surface integrals over floe shapes. The ocean component is given by a two-layer quasi-geostrophic (QG) model that generates eddies from baroclinic instabilities, the so-called Phillips model \cite{vallis2017atmospheric}, in which the cumulative impact of many passing floes on the turbulent eddy field is represented via a quadratic surface drag. The spatial resolution is $128\times128$ gridpoints. The atmospheric  component is taken from a reanalysis product (ERA5) \cite{olauson2018era5, era5_data}, which provides Eulerian wind velocity fields over the observational period. As the wind field exhibits larger-scale features, a coarser spatial resolution of $11\times 11$ gridpoints is used. Note that the focus here is in the MIZ of the Beaufort Sea (see Figure \ref{satellite_fig}). Hence, a double-periodic boundary condition is adopted for simplicity. The potential model error and bias introduced from the various approximations can be mitigated at the statistical forecast stage using the stochastic corrections and Bayesian inference in the framework introduced in the rest of this section. The domain size, as shown in Figure \ref{satellite_fig}, is roughly 600km$\times$600km. Figure \ref{schematic_fig} includes a schematic illustration of the main steps of the framework.

\paragraph{Step 1. Development of low-cost data-driven reduced-order stochastic models.}
The ensemble forecast adopts a probabilistic characterization of the model state and is thus a natural way to predict complex turbulent systems \cite{palmer2019ecmwf, toth1997ensemble, leutbecher2008ensemble}. However, the high dimensionality and nonlinearity of the coupled atmosphere-ice-ocean system makes a single realization of the model forecast very computationally expensive, let alone the forecast of the entire ensemble. Therefore, the first step in this framework is to develop data-driven reduced-order models with the aim to significantly lower the computational cost of the forecast step.

The top section of Figure \ref{schematic_fig} outlines the development of such reduced-order models for the turbulent ocean field. Given a long simulation generated from the original two-layer QG ocean model, the spectral decomposition of the velocity field is used. Most of the energetic modes are concentrated within a circular area centered at $\mathbf{k}=(0,0)$ with a relatively small radius in spectral space $\vert\mathbf{k}\vert\leq K$, where $\mathbf{k}=(k_1,k_2)$ is the spectral index. The reduced-order model is set up to only describe the temporal evolution of the dynamics of this small set of spectral modes. It is expected to retain most of the key features of the original ocean field but significantly lower the computational cost. Yet, given the nonlinearity of the original ocean model, the governing equation of each spectral mode is fully coupled with all other modes, including those omitted in the reduced-order model. To effectively characterize the temporal evolution of each spectral mode in the reduced-order model, a linear stochastic model is developed as a surrogate \cite{gardiner2009stochastic},
\begin{equation}\label{LSM}
    \frac{\d u}{\d t} = (-a + i\omega)u + f + \sigma \dot{W},
\end{equation}
where $u$ is a complex variable for a single spectral mode, $a$ and $\omega$ are the damping and oscillation frequencies, respectively, $f$ is the forcing of the system, $\dot{W}$ is a complex-valued white noise, and $\sigma$ is the amplitude of the noise. In \eqref{LSM}, damping and stochastic noise are adopted to parameterize the contribution of the extremely complicated, nonlinear, and deterministic part of the original governing equation, which leads to a cheaper and more effective way to reproduce the statistical forecast results \cite{majda2016introduction, farrell1993stochastic, berner2017stochastic, branicki2018accuracy, majda2018model}. Independent linear stochastic models are used to characterize the temporal evolution of each mode in spectral space, which nevertheless allows a fully correlated spatial pattern in physical space. Since the QG model generates an incompressible flow field, the spectral representation of the ocean is based on the stream function. On the other hand, a pair of linear stochastic models is utilized to approximate the two-dimensional velocity components associated with each spectral mode of the atmospheric wind field.

\paragraph{Step 2. Systematic model calibration. }
The linear stochastic model in \eqref{LSM} can be calibrated systematically by taking advantage of the analytic formulae for its four fundamental statistics: the mean, the variance, and the real and the imaginary parts of the decorrelation time. The values of these four statistics have a unique one-to-one correspondence with the four parameters $a$, $\omega$, $f$, and $\sigma$. Therefore, once the values of these statistics are computed numerically from the time series of a single spectral mode in the original QG ocean model, these values are plugged into the closed analytic formulae, determining the four parameters in the linear stochastic model associated with that spectral mode.

Rather than simulating the 30,000 modes of the two-layer QG model, setting $K$ to be 11 in step 1 of the framework results in a reduced-order model containing only about 400 modes. This simulation still resembles the full QG system while being much more computationally inexpensive. See panels (a) and (b) of Figure \ref{schematic_fig}. On the other hand, after applying the proposed calibration procedure, a random realization of the time series from the linear stochastic model is also statistically similar to the truth. See panels (c) and (d) of Figure \ref{schematic_fig} for a comparison of mode $(5,5)$. This similarity is essential for an accurate ensemble forecast using the linear stochastic reduced-order models. Finally, the same linear stochastic models are adopted as surrogate models to describe the atmospheric wind field based on the ERA5 reanalysis data.

\paragraph{Step 3. Physics-based dynamical interpolation via nonlinear data assimilation.}
Dynamical interpolation exploits the optimal combination of the ensemble forecast from the model and the information from the partial observations via nonlinear data assimilation. The incorporation of the underlying dynamics sets dynamical interpolation apart from pure curve fitting methods. The basic dynamical interpolation scheme used here is the ensemble Kalman smoother (EnKS), in which an ensemble of model trajectories represents the estimate of the system state. Each ensemble member contains trajectories for all state variables, including the sea ice, the ocean, and the atmosphere. The EnKS provides point estimates through the ensemble mean and quantifies the uncertainty through the ensemble spread. The resulting distribution is called the posterior distribution, which contrasts with the prior distribution solely obtained from the forecast step of the model.

The traditional EnKS contains a straightforward prediction-correction loop in physical space that requires repeatedly integrating the expensive original dynamical model. In contrast, the new method here uses the linear stochastic models to approximate the ocean and atmospheric flow fields, and the prediction-correction procedure alternates between the physical and the spectral spaces. Specifically, the prediction of the ocean and atmospheric flow fields, which involves running the linear stochastic models forward, is implemented in the spectral space. On the other hand, the correction of all the state variables, which applies the Bayesian formula that optimally combines the model and observational information, is carried out in the physical space. Spectral decomposition and flow field reconstruction are adopted after each correction and prediction step, respectively. See panels (g) and (h) of Figure \ref{schematic_fig}. Since only a few spectral modes are involved in the sequential prediction-correction procedure, the computational efficiency is preserved. Note that the DEM sea ice model remains highly nonlinear, which makes the entire dynamical interpolation nonlinear. To further improve the numerical stability and mitigate erroneous spurious correlations, localization and fixed lag strategies are incorporated into the basic version of the EnKS \cite{anderson2012localization, evensen2009data}.

\paragraph{Step 4. Efficient parameter estimation of important physical quantities.}
The main practical challenge of using general dynamical interpolation methods to analyze the sea ice cover is the lack of access to the entire parameter space from a single remote sensing instrument. For example, the thickness of the floes determines the inertia of floe motion and is crucial to the coupled system. To overcome this challenge, an efficient parameter estimation algorithm is embedded into the dynamical interpolation framework. Here, the unobserved physical quantities are treated as the augmented state variables, which are simultaneously estimated with the actual variables of the model state. The uncertainty in the estimated parameters, due to the relatively short Lagrangian trajectories, is also quantified in the algorithm.

\subsection{Setups of the two experiments}
The new dynamical interpolation framework is first applied to a synthetic data experiment and then to the real observation scenario.

The synthetic data experiment uses the two-layer QG ocean model and the reanalysis data for the atmospheric winds to force the ice floes governed by the DEM model. The ice floe shapes, sizes, positions, and orientations are initialized from a library of floes in the Beaufort Gyre MIZ \cite{lopez2021library}, which is generated from optical remote sensing imagery using the Ice Floe Tracker algorithm \citep{lopez2019ice}. The thickness of each floe is randomly drawn from a background distribution \cite{Kwok_2018} and is assumed to be constant during the entire observational period. See the top right panel of Figure \ref{fig:thickness}. Note that the stochastic approximate models are not utilized to generate the synthetic data, rather they are only used to dynamically interpolate the missing floe observations. For the real data experiment, Lagrangian sea ice floe trajectories are obtained using the Ice Floe Tracker algorithm within the study area delineated in Figure \ref{satellite_fig} during the spring-to-summer transition of 2008. The same linear stochastic models that are calibrated for the synthetic data experiment are adopted to carry out the dynamical interpolation. See Table \ref{Table:Setup} for the summary of the models used to generate the true signal and those adopted to implement the dynamical interpolation in the two experiments.

\begin{table} \centering
\begin{tabular}{|c|c|c|c|}
  \hline
        \multicolumn{4}{|c|}{(a) Synthetic data experiment}  \\\hline
        & Atmosphere & Ocean & Sea Ice \\
  Truth & ERA5 reanalysis & Two-layer QG & The known DEM model\\
  Interpolation & Calibrated LSM  & Calibrated LSM & The known DEM model  \\
  \hline
    \multicolumn{4}{|c|}{(b) Real data experiment} \\\hline
   & Atmosphere & Ocean & Sea Ice \\
  Truth & Not needed & Not needed & Satellite observations \\
  Interpolation& Calibrated LSM  & Calibrated LSM & The known DEM model \\\hline
\end{tabular}  \vspace*{5pt}\caption{Summary of the models used for both the synthetic and the real data experiments. In each experiment, the first row ``truth'' stands for the underlying systems that generate the true signal, while the second row ``interpolation'' indicates the model used for dynamical interpolation. The same calibrated linear stochastic model (LSM) is utilized for the real data as for the synthetic data experiments. In the real data case, the true signals of the atmosphere and ocean components are not needed. In the synthetic data case, the true atmosphere and ocean models are used to drive the DEM model to generate the observed floe trajectories and angular displacements.} \label{Table:Setup}
\end{table}

The floe locations and angular displacements are the only observational information in the dynamical interpolation for the coupled atmosphere-ice-ocean system. These two quantities are obtained from the satellite images at a frequency of roughly every 24 hours. The observational uncertainty, which is used in the dynamical interpolation algorithm, is set to be 0.25 km and 5 degrees, respectively.

Both experiments contain 38 floe trajectories of various lengths. Excluding the first and the last point in each floe trajectory, there are in total 164 remaining candidate observational points for the 38 trajectories. These 164 candidates are randomly divided into four sets, where each set contains 41 data points. Then four independent dynamical interpolation simulations are carried out. In each simulation, the 41 candidate observations in the corresponding set are artificially removed as the missing observations. Note that the missing observations referenced in the real data experiment are not the actual missing ones in the satellite images obscured by clouds, but are rather the artificially removed ones. Such a setup guarantees the true values of these missing floes are known and therefore it allows the qualitative study of the accuracy of the dynamical interpolation. Nevertheless, this one-third ratio between the number of observed and missing floe observations mimics the real-world situation in the MIZ during the boreal summer.

\begin{figure}[h]
\centering
\includegraphics[width=\textwidth]{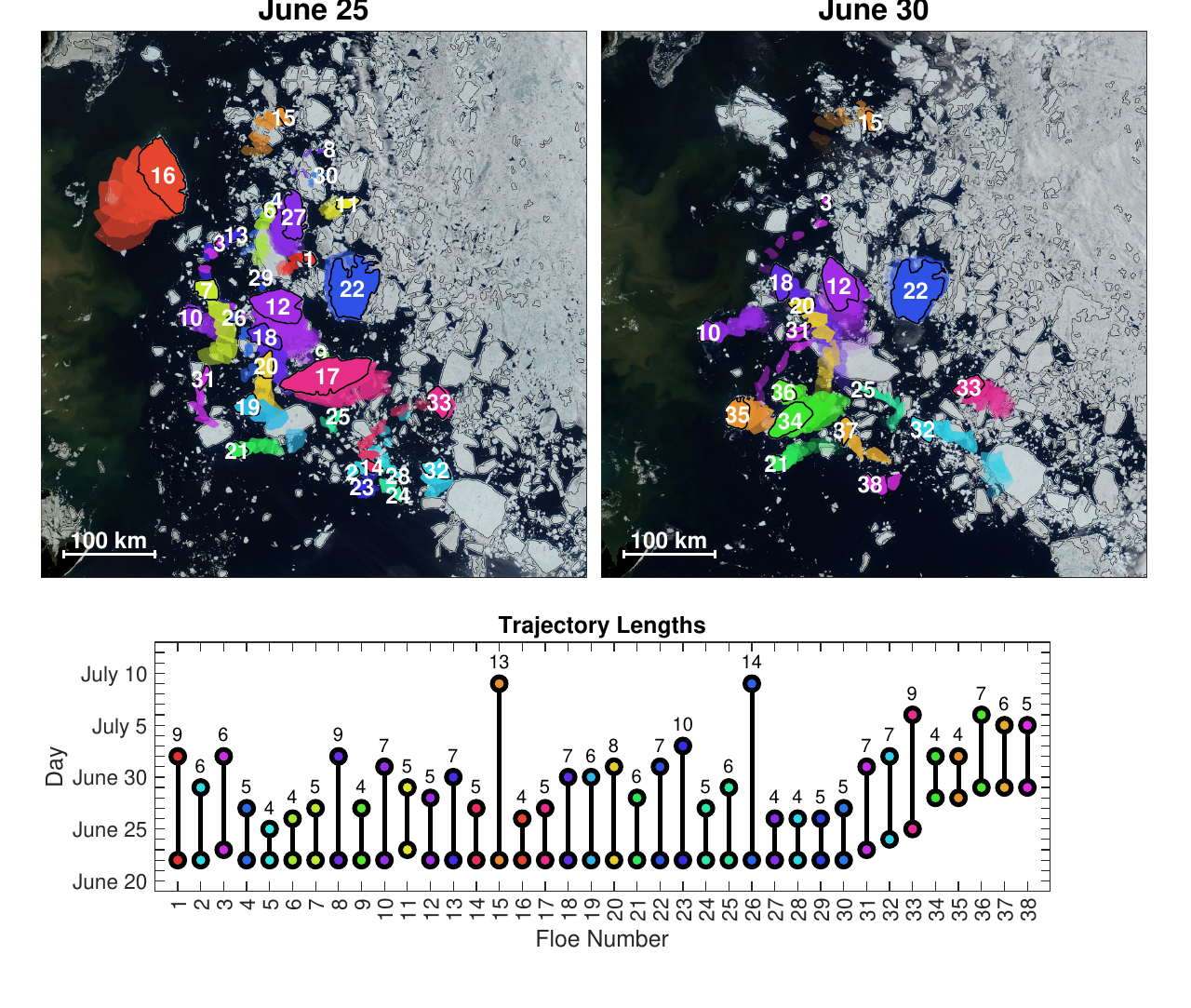}
\caption{Sea ice floe trajectories retrieved from optical satellite remote sensing imagery. MODIS True Color images (downloaded from the NASA Worldview application) acquired on 25.06.2020 and 30.06.2020 are displayed in a WGS 84/NSDIC Sea Ice Polar Stereographic North 70'' N Projection, on top of which retrieved ice floe trajectories are displayed in color. In both images, the evolution of floe positions is represented as a shift in opacity from transparent to opaque objects. Final floe positions are marked using black contour lines. Note that the recovered floe trajectories have different lengths and periods. Information regarding the acquisition period of each floe trajectory is shown in the bar plot underneath. Only a sub-set of the 38 non-interacting floes used in this study are shown for clarity. }\label{overlay_fig}
\end{figure}

Figure \ref{overlay_fig} displays the 38 sea ice floe trajectories in the real data experiment, which are retrieved from satellite remote sensing imagery using the Ice Floe Tracker algorithm during the spring-to-summer transition of 2008. Each floe trajectory is represented by the transition from fully transparent to opaque with each floe assigned a specific color. The index in these floes corresponds to those in Figures \ref{fig:interpolation} and \ref{many_interpolations_fig}.

\subsection{Results of the synthetic data experiment}

\begin{figure}[h]
\centering
\includegraphics[width=\textwidth]{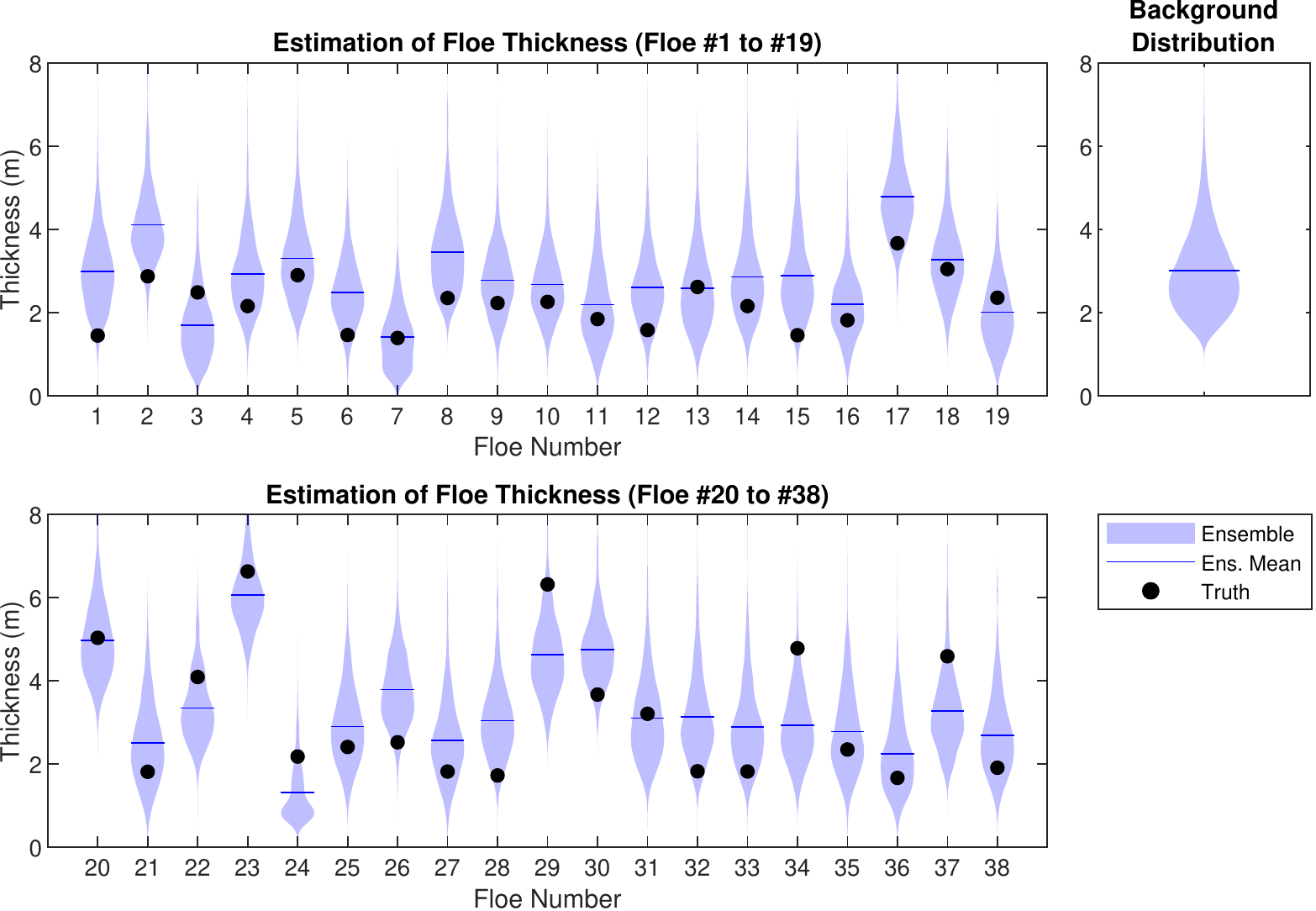}
\caption{Parameter estimation of sea ice thickness in the synthetic data experiment. Left panel: The black dots and solid lines indicate the truth and the ensemble mean estimate of each sea ice floe, respectively, while the shaded area in the violin plot indicates the estimated non-Gaussian PDF formed by ensembles. Right panel: The background sea ice thickness distribution. The true value of the thickness for each sea ice floe is randomly drawn from such a distribution.  It is also used as the initial distribution in the parameter estimation algorithm. }
\label{fig:thickness}
\end{figure}

Figure \ref{fig:thickness} illustrates the parameter estimation  of the floe thicknesses from the dynamical interpolation. To quantify the uncertainty in the estimated thickness of each floe, the posterior distribution characterized by the ensemble members is also included via a violin plot.  The results shown here are calculated using all the ensemble members from the four simulations, but different simulations lead to similar distributions for all the floes. The estimations are overall reasonably accurate, especially given such a small number of observations within a large domain. Particularly, the truth of all the 38 floes is consistently covered by the posterior distribution. In addition, for two-thirds of the floes, the true thickness value lies in a high likelihood region of the distribution within one standard deviation from the mean. Note that the error in the thickness estimation can offset the error from recovering the atmosphere and ocean flow fields, and therefore the overall interpolation results remain accurate, as will be seen below. It is worthwhile to highlight that both the background thickness distribution, from which the true thickness values are drawn, and the estimated posterior distributions, exhibit strongly fat-tailed non-Gaussian behavior, as is clear in the violin plot. These non-Gaussian features are the unique outcome of the highly nonlinear dynamics of the sea ice floes. In addition, because the ensemble is transformed from the prior distribution to the posterior during each update of the nonlinear EnKS, rather than resampled, these important non-Gaussian features can be preserved through the ensemble update (see the Methods section.) By contrast, simply considering the ensemble mean and standard deviation would underestimate the likelihood of large floe thicknesses while simultaneously overestimating the likelihood of small thicknesses. Notice that such a non-Gaussian feature is found in all model variables, but is especially illustrated by thickness estimation. These findings imply the necessity of incorporating both the nonlinear sea ice dynamics and the nonlinear data assimilation scheme into the dynamical interpolation framework.

\begin{figure}[h]
\centering
\includegraphics[width=\textwidth]{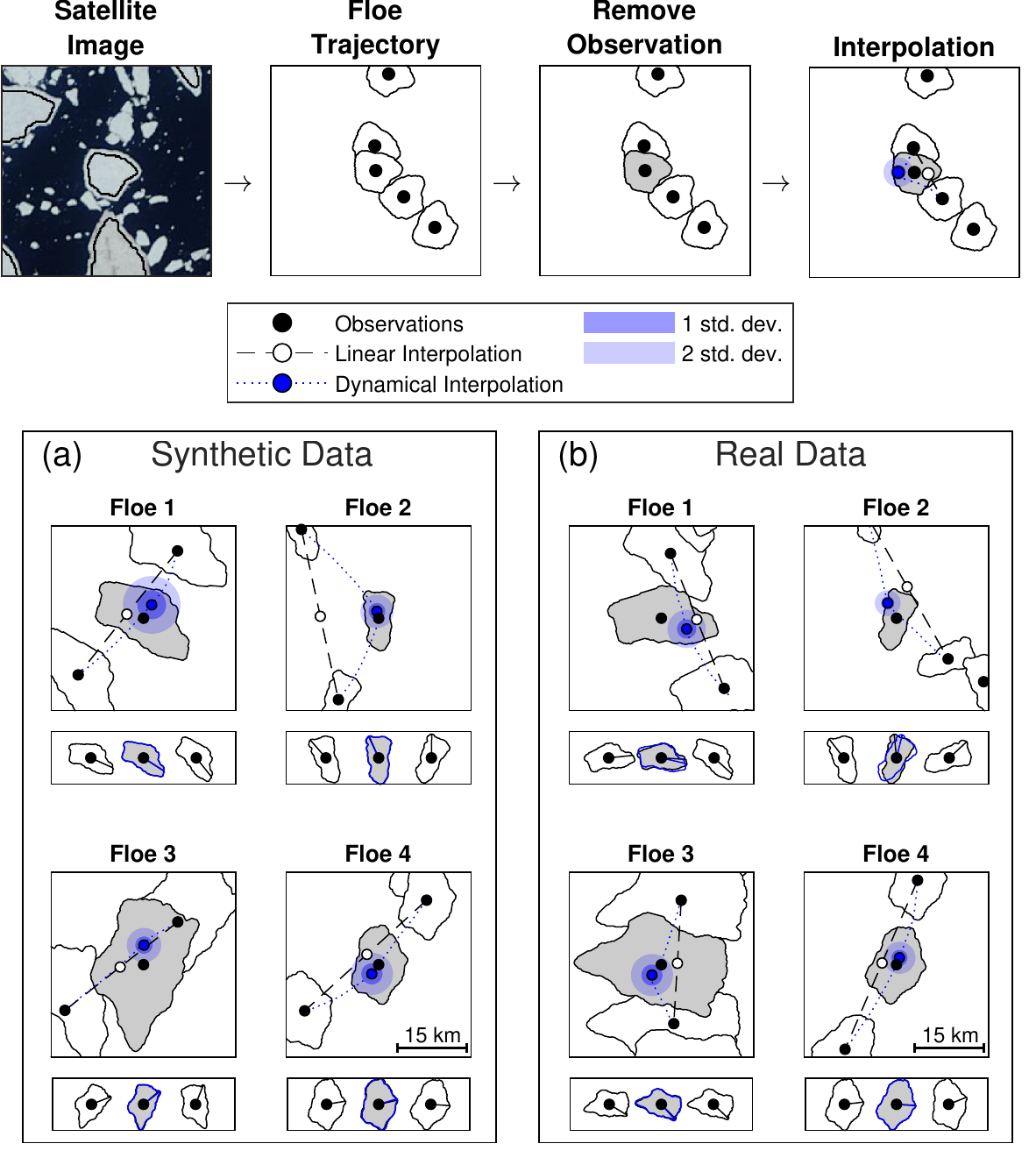}
\caption{Comparison of recovering the missing observations using linear and dynamical interpolation schemes. The top panel illustrates the procedure of performing the interpolation experiments. A floe trajectory is first retrieved from the satellite imagery. Next, the observed floe on a specific day is artificially removed. The linear/dynamical interpolation framework is applied to recover this artificially removed observation. In the bottom part, panels (a) and (b) show the results from the synthetic and the real data experiments, respectively. In each panel, the top part shows the interpolated floe locations while the bottom part shows the interpolated angular displacement. Since the floes in the synthetic data experiment are taken from the library of sea ice floe observations, floes with the same index in the two experiments are identical (i.e., shapes and sizes are retained). In addition to the ensemble mean estimate presented by the blue marker, the uncertainty resulting from the dynamical interpolation is provided by the shaded areas. For the illustration purpose, only the two-dimensional Gaussian confidence interval is used to characterize the uncertainty in the dynamical interpolation. }
\label{fig:interpolation}
\end{figure}

Panel (a) in Figure \ref{fig:interpolation} compares linear and dynamical interpolation for recovering the floe location and angular displacement. The ensemble mean estimate using the dynamical interpolation almost always outperforms linear interpolation in recovering the floe locations. Specifically, the absolute error using the linear interpolation is nearly three times as large as that using the dynamical interpolation. The linear interpolation also, by design, completely fails to recover the curvature and the nonlinear evolution of the floe trajectories. In contrast, the dynamical interpolation accurately captures these important physical features. In addition to the point estimate using the ensemble average, the ensemble provides the quantification of the estimated uncertainty. The uncertainty overall remains at a relatively low level, indicating the confidence of the posterior mean estimate. Among all the 164 recovered missing observations, roughly 80\% of the true observations fall within two standard deviations around the ensemble mean estimate. This implies the accuracy and robustness of the dynamical interpolation. In addition, the recovery of the angular displacement using the dynamical interpolation is quite accurate.

\begin{figure}[h]
\centering
\includegraphics[width=10cm]{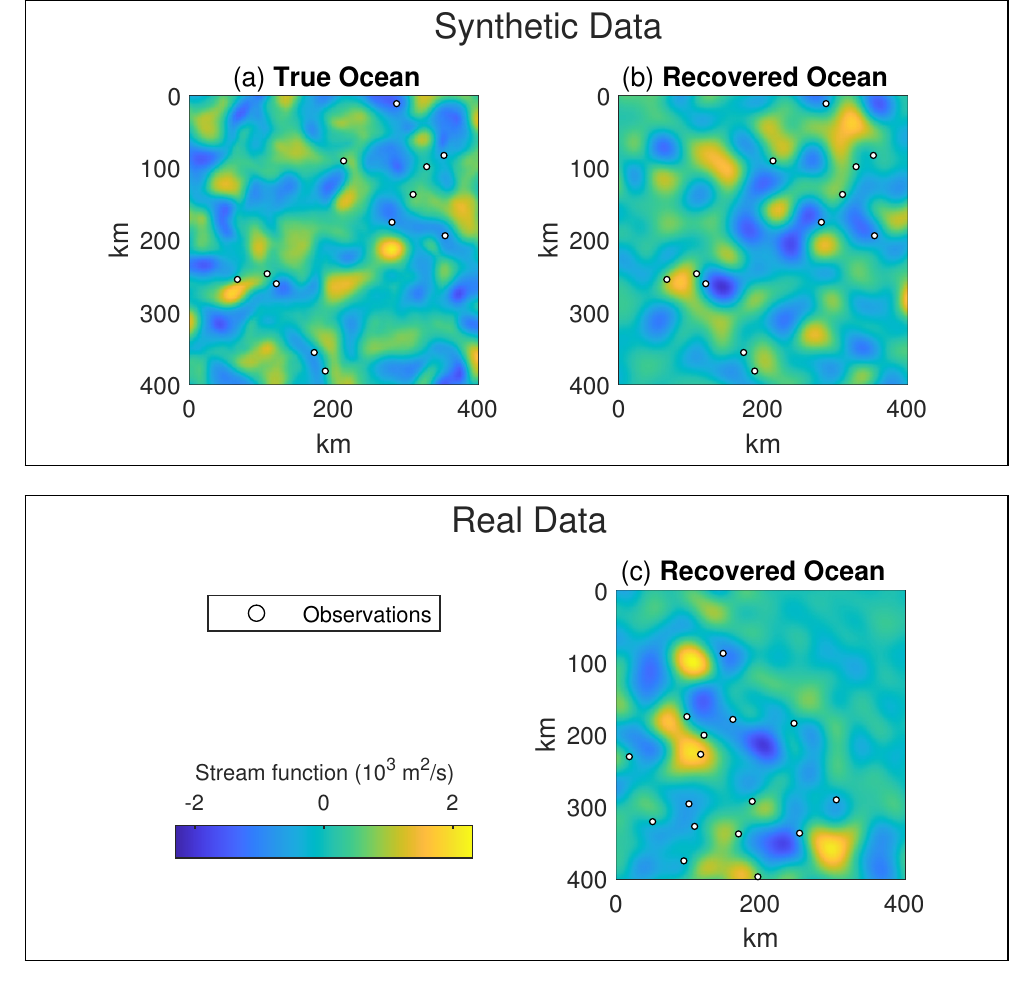}
\caption{The recovered ocean flow field represented by the stream functions utilizing the dynamical interpolation. The top panel shows the truth and the recovered ocean field in the synthetic data experiment while the bottom panel shows the recovered ocean field of the real data. Since the primary focus is to resolve regions close to the ice edge, a 400km$\times$400km domain within the original 600km$\times$600km area and having the same domain center is presented. The results shown here are on a specific day in the middle of the study period. For the real data, it is July 1. The error in recovering the ocean field remains in a similar level on other days. The white dots mark the locations of the floes. There are in total 17 floes inside the 400km$\times$400km domain for both cases.}\label{ocean_fig}
\end{figure}

Panel (a) of Figure \ref{ocean_fig} illustrates the recovered ocean field represented by the stream functions utilizing the dynamical interpolation. The result shown here is on a specific day in the middle of the entire time period. The accuracy in recovering the ocean field remains in a similar level on other days. The overall pattern correlation between the truth and the recovered ocean field is around $0.5$. Nevertheless, given the fact that there are only $17$ floes inside this large domain on this day, the skill of recovering the ocean field is already significant. In particular, the ocean eddies are recovered quite reasonably in the areas, where the observed floes are concentrated. The SI includes more sensitivity analysis, which shows the improvement of the recovered ocean field if the density of observed floes increases.

\clearpage
\subsection{Results of the real data experiment}\label{Sec:RealData}

\begin{figure}[h]
\centering
\includegraphics[width=\textwidth]{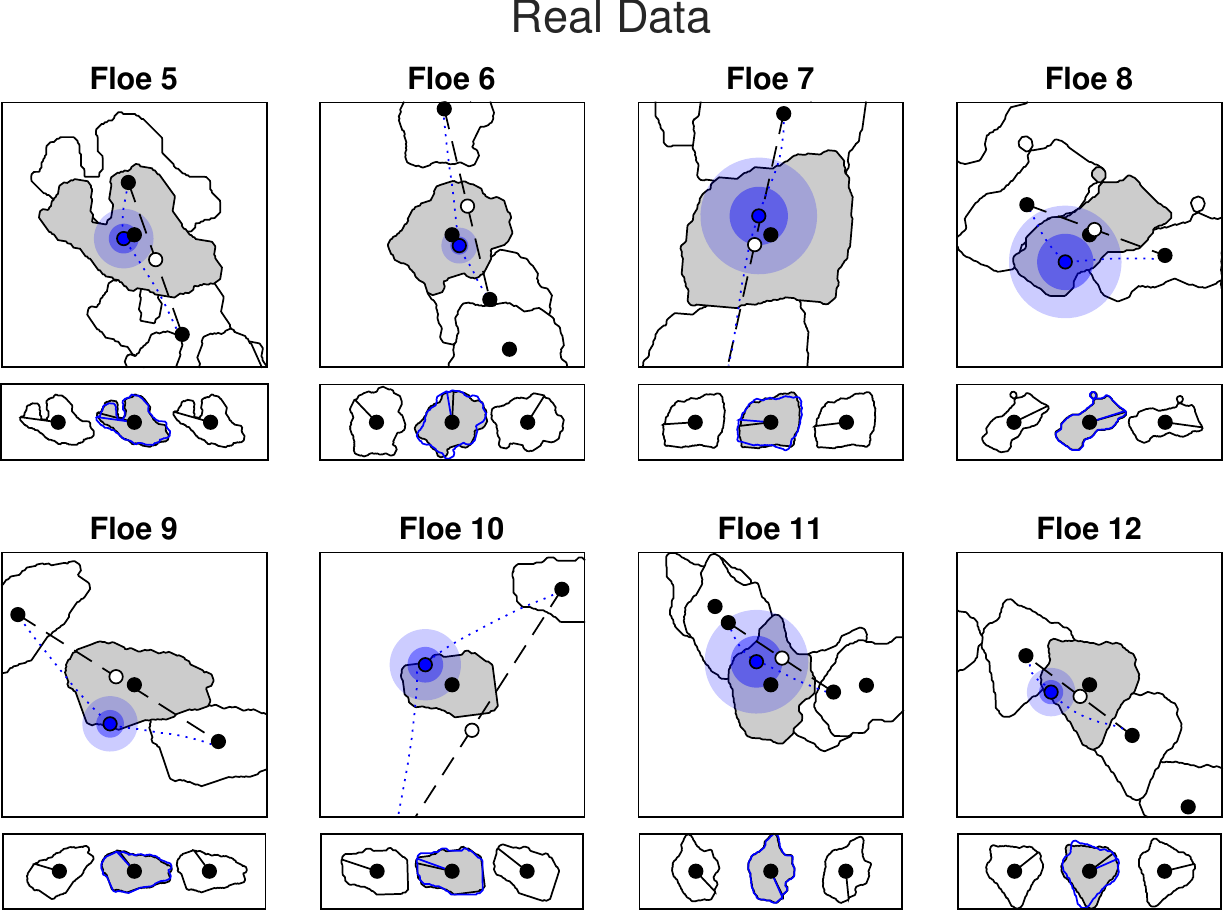}
\caption{Additional results for the real data experiment, similar to those in Panel (b) of Figure \ref{fig:interpolation}. }\label{many_interpolations_fig}
\end{figure}

Panel (b) of Figure \ref{fig:interpolation} includes  four cases of the recovered missing floes on the real data set. Similar to the conclusion from the synthetic data experiment, the dynamical interpolation being applied to the real data set also shows significant advantages over the linear interpolation in the sense that the error in the ensemble mean is overall much smaller and the uncertainty can be systematically quantified. Comparing with the analogs from the synthetic data experiment in Panel (a), the accuracy of the results in the real data test remains comparable. Figure \ref{many_interpolations_fig} includes additional case studies of the recovered missing floe trajectories from the real data experiment. Again, the dynamical interpolation provides reasonable results in most of the cases.
Panel (b) of Figure \ref{ocean_fig} displays the recovered ocean field on July 1, 2008. Although there is no true solution for the validation of the point-wise recovery skill, the overall flow amplitudes as well as the number and the size of the eddies in the recovered ocean field all look reasonable. One interesting finding is that the recovered ocean field in the north-east corner of the domain is nearly zero due to high uncertainty, which corresponds to the area beneath the large piece of the ice cover shown in Figure \ref{overlay_fig}.

\begin{figure}[h]
\centering
\includegraphics[width=\textwidth]{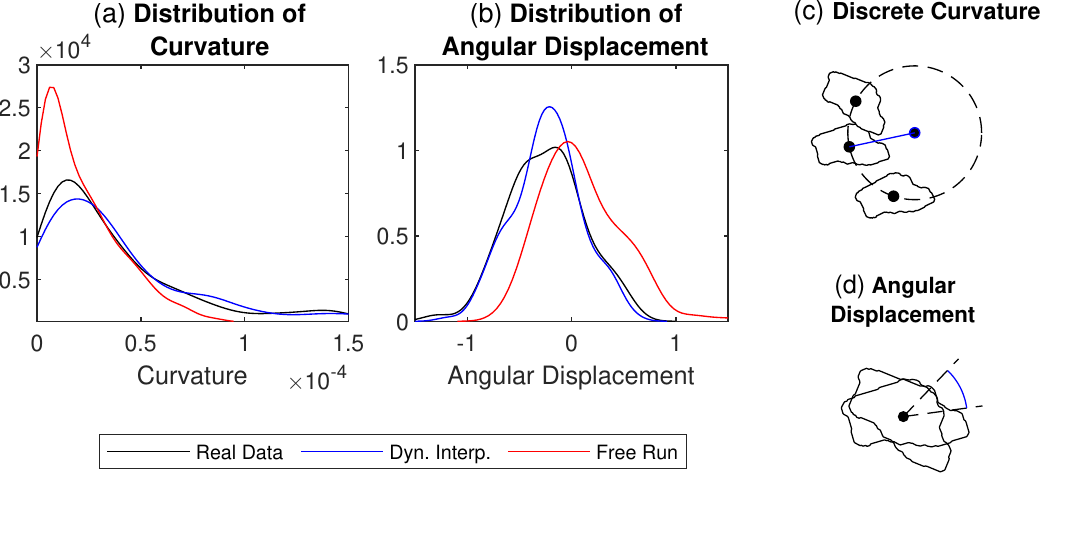}
\caption{Comparison of the recovered properties using different interpolation methods. Panel (a): comparison of the distribution of the curvature of the recovered trajectories. Panel (b): comparison of the distribution of the angular displacement of the recovered trajectories. Panels (c)--(d): Schematic illustrations of the definitions of the discrete curvature and angular displacement used to compute the distributions in Panels (a)--(b).  }
\label{fig:trajectory}
\end{figure}

Figure \ref{fig:trajectory} compares the physical properties of the recovered ice floes between the real observations, the dynamically interpolated data, and the direct model simulation.
Since the data set consists of discrete observations, the two metrics used are the discrete curvature and the daily angular displacement. The former is calculated using the circumscribing circle of each trio of observations while the latter is obtained by taking the difference in angle between two consecutive observations. The results using the linear interpolation are omitted here, as the linear interpolation fails to provide any useful information of these two physical quantities.
Panel (a) shows that the curvature of the floe trajectories from the direct model simulation is severely underestimated, which is a natural outcome of the model error. In contrast, the data resulting from the dynamical interpolation succeeds in reproducing the non-Gaussian distribution of the observed truth with a one-sided fat tail.
Next, with respect to the angular displacement, as is shown in Panel (b), the real data set has a negative bias due to the influence of the Beaufort Gyre, something which is not reflected in the direct model simulation that is again due to the model error.
Nevertheless, such a bias in the direct model simulation is almost fully corrected in the dynamically interpolated data with the help of the partial observations.
These results indicate the importance of utilizing both the observations and a suitable model in the dynamical interpolation, as the model provides at least partially the access to the crucial underlying nonlinear dynamical information while the observations can largely reduce the  biases from the model forecast.

\section{Discussion}
Model error is inevitable when studying complex systems. In the dynamical interpolation framework developed here, the sea ice DEM model remains highly nonlinear while the ocean and atmospheric components are effectively approximated by linear stochastic models. Indeed, a large error will appear if these linear stochastic models are used to study the dynamics associated with the ocean and atmospheric fields. Nevertheless, for the purpose of dynamical interpolation, the information needed from the model is merely some prior knowledge of the short-range statistical forecast of these fields, which are usually quite accurate due to the fact that these linear stochastic models are carefully calibrated.

The reduced-order models in the proposed framework are not limited to linear stochastic models. If the time series of the underlying flow fields exhibit strong non-Gaussian features, then suitable nonlinear or non-Gaussian surrogate models can be easily incorporated \cite{averina1988numerical, chen2018conditional, gershgorin2010improving}. One such simple candidate is a family of linear models with multiplicative noise. Another important point is that the sea ice dynamics within the scales studied here are predominantly nonlinear. The governing equations are well understood and are crucial in the dynamical interpolation, given that the directly observed variables are sea ice floe trajectories. The strong nonlinearity in sea ice dynamics is also more deterministic and less turbulent than the atmosphere or the ocean. Therefore, linear stochastic models are not appropriate for approximating the fully nonlinear behavior of sea ice. Since the degree of freedom in characterizing the floe trajectories is much lower than the governing equations of the atmospheric and oceanic velocity fields, the nonlinear floe dynamics are explicitly incorporated into the dynamical interpolation framework.

It is also worth highlighting the importance of the prior physical model of the ocean and the prior time series of the atmosphere, which significantly facilitate the calibration of the linear stochastic models. In the absence of a suitable prior model or data for the ocean and the atmosphere, the calibration of the reduced-order stochastic surrogate models requires a more complicated iterative expectation-maximization procedure \cite{chen2020learning}. In other words, the dynamical interpolation, the parameter estimation of the thickness, and the uncertainty quantification of the oceanic and atmospheric flow fields have to be carried out simultaneously with the floe trajectories providing the only available information. Such an iterative approach often requires an extensive observational database to ensure the accuracy of the dynamical interpolation scheme.

Another crucial point is the quantification of the uncertainty, particularly in the presence of the model error, the small number of observations, and the implications of turbulent systems. Uncertainty quantification is not available by applying direct curve fitting methods but rather a unique feature of the dynamical interpolation framework. In this study, the properties of sea ice exhibit various non-Gaussian features such as the non-symmetry in the distribution of the angular displacement,  strong skewness, and fat tails with extreme events in the distributions of the curvature and the thickness. These non-Gaussian features have been shown to be crucial in understanding the sea ice dynamics \cite{toppaladoddi2015theory, moon2017stochastic}. Hence, to assess uncertainty, the attributes of the entire distribution are considered.

Finally, the framework developed here has several unique implications for improving our understating of Earth system science.
First, new-generation climate models that accurately represent sea ice dynamics at the floe scale will require validation against Lagrangian observations of sea ice floes at high and moderate resolutions. The point-estimate recovery of missing observations and the associated estimates of the uncertainty mitigate some of the issues of Lagrangian optical remote sensing observations. The methodology presented here is also easily adaptable to analyze the output of other instruments. In this sense, it is expected that the continuous trajectories stemming from the nonlinear data assimilation can be used in more sophisticated deep-learning models for calibration and training.
Second, the proposed framework allows for accurate parameter estimation of unobserved variables at unprecedented scales in a Lagrangian setting. For example, the retrieval of sea ice thickness is outlined here, which is a crucial variable in understanding the evolution of the sea ice cover in response to a changing climate.
Lastly, data assimilation provides the key missing piece for understanding ocean transport and mixing processes at high latitudes. Small-scale eddies have an important role in transferring energy to larger-scale structures via an inverse cascade of energy and are thus hypothesized to be the missing energy source to close the ocean energy budget. High-resolution numerical simulations have highlighted their contribution to nutrient redistribution, oxygen transport, and biogeochemical processes. However, they are hard to observe due to the lack of resolution of space-borne sensors and the sparsity of in situ instruments. Given the recently demonstrated connections between the rotation rate of sea ice floes and eddies with sub-surface expression in the western Arctic Ocean, it is anticipated that this method can be used to understand fundamental processes of ocean turbulence at small-to-moderate scales.

\section{Methods}

\subsection{The coupled atmosphere-ice-ocean system}
\subsubsection{The DEM model}

The sea ice floes are described by a DEM model. The floes can have arbitrary 2D shapes with their movements and rotations determined by the surface integrals of the ocean and wind velocities over these shapes. Both the full QG model and reanalysis data or and the approximate stochastic models can be used to drive this ice floe model.
In the DEM model utilized here, the shape and thickness for each floe are assumed to be unchanging over time. Since the non-interacting floes are the primary focus of this work, the contact forces are not included in the model presented here, which greatly reduces the computational cost.

The dynamics of a single ice floe is described as follows \cite{manucharyan2022spinning, chen2022efficient}. Let $\mathbf{X}_{\mathrm{ice}} = (x_{\mathrm{ice}}, y_{\mathrm{ice}})$ be the centroid of the floe and $\Omega$ its the angular displacement about the centroid. Also denote $\mathbf{V}_{\mathrm{ice}} = (u_{\mathrm{ice}}, v_{\mathrm{ice}})$ to be the velocity of the floe and $\omega$ is the angular velocity.
Then ice floe-ocean interactions are calculated using surface integrals over the area of the floe:
\begin{align}
\dot{\mathbf{X}}_{\mathrm{ice}} =& \mathbf{V}_{\mathrm{ice}} \\
\dot{\Omega} =& \omega \\
m\dot{\mathbf{V}}_{\mathrm{ice}} =& \iint_A \mathbf{F}_{\mathrm{total}} \, \mathrm{d}A \\
I\dot{\omega} =& \iint_A \tau\, \mathrm{d}A
\end{align}
where $A$ is the area of the floe. $\mathbf{F}_\mathrm{total}$ is the total force on the ice floe induced by the ocean, atmosphere, and other sources. $\tau$ is the resulting torque, calculated from the force.

The total force on the floe by the ocean consists of ocean drag, atmosphere forcing, Coriolis force, and the pressure gradient
\begin{equation}
\mathbf{F}_{\mathrm{total}} = \mathbf{F}_{\mathrm{ocn}} + \mathbf{F}_{\mathrm{atm}} + \mathbf{F}_{\mathrm{Coriolis}} + \mathbf{F}_{\mathrm{pressure}}.
\end{equation}
To incorporate the ocean turning angle, define the rotation matrix $\mathbf{R}_\theta$ as
\begin{equation}
\mathbf{R}_\theta = \left(\begin{matrix}\cos(\theta) & -\sin(\theta) \\ \sin(\theta) & \cos(\theta)\end{matrix}\right).
\end{equation}
The force induced by the ocean drag at the point $\mathbf{X}_\mathrm{ocn}$ within $A$ is given by
\begin{equation}
\mathbf{F}_{\mathrm{ocn}} = \rho_{\mathrm{ocn}} C_{\mathrm{ocn}} \|\bm{V}_{\mathrm{ocn}}-\bm{V}_{\mathrm{ice}}\|\mathbf{R}_\theta(\mathbf{V}_{\mathrm{ocn}}-\mathbf{V}_{\mathrm{ice}})
\end{equation}
where $\theta$ is the fixed ocean turning angle.
Similarly force induced by the ocean drag at the point $\mathbf{X}_\mathrm{atm}$ within $A$ is given by
\begin{equation}
\mathbf{F}_{\mathrm{atm}} = \rho_{\mathrm{atm}} C_{\mathrm{atm}} \|\bm{V}_{\mathrm{atm}}-\bm{V}_{\mathrm{ice}}\|(\mathbf{V}_{\mathrm{atm}}-\mathbf{V}_{\mathrm{ice}})
\end{equation}
where $\theta$ is the fixed ocean turning angle.
The Coriolis force is constant over the area of the of the floe and is given by
\begin{equation}
\mathbf{F}_\mathrm{Coriolis} = \rho_\mathrm{ice} f_c L_{\mathrm{ice}} \mathbf{R}_{-\pi/2} \mathbf{V}_\mathrm{ice}
\end{equation}
The force induced by the pressure gradient is similar but depends on the ocean velocity and so varies over $A$
\begin{equation}
\mathbf{F}_\mathrm{pressure} = \rho_\mathrm{ice} f_c L_{\mathrm{ice}} \mathbf{R}_{\pi/2} \mathbf{V}_\mathrm{ocn}.
\end{equation}
The torque induced on the floe at the grid point $\mathbf{X}_\mathrm{torque}$ is given by
\begin{equation}
\tau = (\mathbf{X}_\mathrm{torque} - \mathbf{X}_{\mathrm{ice}}) \times \mathbf{F}_{\mathrm{total}} = (x_\mathrm{torque} - x_\mathrm{ice})F_y-(y_\mathrm{torque} - y_\mathrm{ice})F_x
\end{equation}
where $\mathbf{F}_{\mathrm{total}} = (F_x, F_y)$ are the components of the total force.

\subsubsection{The two-layer QG model}
The ocean model is a two-layer Quasi-Geostrophic (QG) model with periodic boundary conditions on a square domain. The ocean state is characterized by the stream functions $\psi_i(x, y)$ and potential vorticities (PV) $q_i(x, y)$ of each layer $i = 1, 2$. The level curves of the stream function, $\psi_i$, correspond to streamlines of the velocity field, which guarantees an incompressible flow. The ocean velocity field for each layer can thus be calculated as
\begin{equation}
(u_i, v_i) = \left(-\frac{\partial \psi_i}{\partial y}, \frac{\partial \psi_i}{\partial x}\right), \quad i = 1, 2. \label{eqn:vel}
\end{equation}

The formulation of the QG equations follows the version in \cite{arbic2004baroclinically}. The PDEs which govern the time evolution of $\psi_i$ and $q_i$ are as follows:
\begin{align}
\frac{\partial q_1}{\partial t} + \overline{u_1} \frac{\partial q_1}{\partial x}
+ \frac{\partial \overline{q}_1}{\partial y}\frac{\partial \psi_1}{\partial x} + J(\psi_1, q_1) =& \mathrm{ssd} \\
\frac{\partial q_2}{\partial t} + \overline{u_2} \frac{\partial q_2}{\partial x}
+ \frac{\partial \overline{q_2}}{\partial y} \frac{\partial \psi_2}{\partial x} + J(\psi_2, q_2) =& -R_2 \nabla^2 \psi_2 + \mathrm{ssd}.
\end{align}
Here ``ssd'' represents small-scale dissipation, which are higher-order derivative terms that are ignored. $J$ is the Jacobian
\begin{equation}
J(\psi, q) = \frac{\partial \psi}{\partial x} \frac{\partial q}{\partial y} - \frac{\partial \psi}{\partial y} \frac{\partial q}{\partial x}.
\end{equation}
The stream functions further satisfy
\begin{align}
q_1 =& \nabla^2 \psi_1 + \frac{(\psi_2 - \psi_1)}{(1 + \delta) L_d^2} & q_2 =& \nabla^2 \psi_2 + \frac{\delta(\psi_1 - \psi_2)}{(1 + \delta) L_d^2}.
\end{align}
where $\delta = H_1/H_2$, $H_i$ is the depth of each layer, and $L_d$ is the deformation radius.

$\partial \overline{q}_1/\partial y$ and $\partial \overline{q}_2/\partial y$, despite the notation, are parameters representing the mean PV gradients for each layer and are given by
\begin{align}
\frac{\partial \overline{q_1}}{\partial y} =& \frac{\overline{u_1} - \overline{u_2}}{(1 + \delta)L_d^2} & \frac{\partial \overline{q_2}}{\partial y} =& \frac{\delta (\overline{u_2} - \overline{u_1})}{(1 + \delta) L_d^2}
\end{align}
where $\overline{u_1}$ and $\overline{u_2}$ are the mean ocean velocities. The final parameter, $R_2$, is the  decay  rate  of  the  barotropic  mode
\begin{equation}
R_2 = \frac{f_0 d_{\mathrm{Ekman}}}{2H_2}
\end{equation}
where $f_0$ is the Coriolis parameter and $d_{\mathrm{Ekman}}$ is the bottom boundary layer thickness. Note that in this formulation we use a constant Coriolis force throughout the domain.

Table 2 summarizes the parameters in the DEM and two-layer QG models.
\begin{table}
    \centering
    \begin{tabular}{cc}
        Parameter & Value \\ \hline
        Ocean density & $\rho_{\text{ocn}} =1027$kg/m$^3$  \\
			Ice density &$\rho_{\text{ice}}=920$kg/m$^3$ \\
			Air density & $\rho_{\text{atm}}=1.2$kg/m$^3$  \\
			Ocean drag coefficient & $c_{\text{ocn}}=5.5\times 10^{-3}$\\
			Atmosphere drag coefficient & $c_{\text{atm}}=1.6\times 10^{-3}$\\
			Coriolis coefficient & $f_c=1.4\times 10^{-4}$\\
			Top layer mean ocean velocity & $\overline{u_1}=2.58$km/day\\
			Bottom layer mean ocean velocity & $\overline{u_2}=1.032$km/day\\
			Top layer mean potential vorticity&$\frac{\partial \overline{q_1}}{\partial y}=0.0265$km$^{-1}$day$^{-1}$\\
			Bottom layer mean potential vorticity&$\frac{\partial \overline{q_2}}{\partial x}=-0.0212$km$^{-1}$day$^{-1}$\\
			Coriolis parameter&$f_c=12$day$^{-1}$\\
			Coupling parameter &$R_1=6.9\times10^{-5}$km$^{-1}$\\
			Decay rate of the barotropic mode&$R_2=1$day$^{-1}$\\
			Deformation radius	&$L_d=5.7$km\\
			Ratio of upper-to lower-layer depth&$\delta=0.8$\\
			Turning angle of the ocean & $\theta=\pi/9$\\
            Ensemble size  &  600\\
            Localization radius & 200 km\\
            Observational noise in location & 250 m\\
            Observational noise in angular displacement & $5^\circ$\\
    \end{tabular}
    \caption{Parameters in the DEM, the two-layer QG models and the EnKS.}
    \label{tab:parameters}
\end{table}

\subsubsection{The atmospheric wind velocity data}
The fifth generation ECMWF reanalysis data product (ERA5) \citep{olauson2018era5, era5_data} for the global climate and weather is implemented for describing the atmospheric wind that is used to calibrate the atmospheric component of the linear stochastic models.

\subsection{Sea ice floe observations and the processing of satellite images}

Remote sensing measurements were retrieved from Moderate Resolution Imaging Spectroradiometer (MODIS) optical imagery (Level 1B 250M). The data is open-access through the Earth Observing Sysetm Data and Information System (EOSDIS) Worldview platform (\url{https://worldview.earthdata.nasa.gov}). In summary, both Corrected Reflectance True and False Color images were pre-processed to reduce the imprint of atmospheric noise allowing the segmentation of sea ice floes ranging from 4 to 75 km in length scale as individual objects. Ice floes were then tracked in a three-stage process involving comparing geometrical parameters in successive images, finding potential matches, and selecting the best candidates based on the assessment of a similarity metric and surface area differences. The reader is referred to \citep{lopez2019ice} for a detailed description of the pre-processing, segmentation, and tracking routines.

\subsection{Calibration of Stochastic Forecast Models}

Statistically accurate  stochastic models are used for the ocean and atmosphere components of the forecast model.
These models can be systematically calibrated based on a long simulation of the two-layer QG model in the case of the ocean component and the reanalysis data set in the case of the atmosphere component. In both cases, the system state is represented in spectral space and the evolution of each spectral mode is governed by a linear stochastic model \eqref{LSM}.
Only the modes with a wave number less than a certain radius are kept: $\vert k\vert \leq 11$ in the case of the ocean and $\vert k\vert \leq 5$ in the case of the atmosphere for a total of 337 and 81 modes respectively.

Recall the linear stochastic model in \eqref{LSM}, which is also known as the complex Ornstein–Uhlenbeck (OU) process \cite{gardiner2009stochastic}. In \eqref{LSM},
 $a$, $\omega$,  and $\sigma$ are real-valued parameters with $a, \sigma > 0$, $f$ is a complex-valued parameter, and $\dot{W}$ is a
complex-valued white noise.
The equilibrium distribution of this OU process is Gaussian and its mean and variance are given in terms of the model parameters:
\begin{align}
\bar{u} =& \frac{f}{a - i \omega} & \operatorname{Var}(u) =& \frac{\sigma^2}{2a}.
\end{align}
The decorrelation time is defined as
\begin{equation}
    T = \int_0^\infty \frac{\mathbb{E}\left[(u(t) - \overline{u})(u(t + \tau) - \overline{u})^\ast\right]}{\operatorname{Var}(u)} \, dt
\end{equation}
and is also given in terms of the model parameters.
\begin{align}
T =& \frac{1}{a - i \omega}.
\end{align}
Using these equation for the equilibrium mean, variance, and decorrelation time, the four parameters of the OU process, $d$, $\omega$, $f$, and $\sigma$, can be written explicitly in terms of these equilibrium statistics as in
\begin{align}
d =& \operatorname{Re}\left[\frac{1}{T}\right]
&
\omega =& -\operatorname{Im}\left[\frac{1}{T}\right]
&
f =& \frac{\bar{u}}{T}
&
\sigma =& \sqrt{2 \operatorname{Var}(u)\operatorname{Re}\left[\frac{1}{T}\right]}.
\end{align}
Using these formulae for the model parameters, an OU process can be fit to match a given set of equilibrium statistics.
In the case of the ocean model, an independent OU process is fit to each spectral mode of the stream function using the mean, variance, and decorrelation time of a long QG model simulation.
For the atmosphere model, a pair of independent OU processes is fit to each spectral mode of the two-dimensional velocity field using the statistics of the velocity field from the ERA5 data set.

\subsection{Ensemble Update}

The ensemble Kalman smoother (EnKS) \cite{evensen2009data} represents the system state with an ensemble of model trajectories.
This ensemble is iteratively forecast and updated based on a set of observations, which are processed sequentially in time.
Let $K$ denote the number of days of observations and denote the time of each day of observations by $t_k$ for $k = 1, \dots, K$.
Denote the $M$-dimensional system state at time $t$ by $\boldsymbol{\psi}(t)$ for $t_1 \leq t \leq t_K$.
Then define the vector $\boldsymbol{d}_k$ of observed floe locations and orientations at time $t_k$ by
\begin{equation}
\boldsymbol{d}_k = \mathcal{M}_k\left[\boldsymbol{\psi}(t_k)\right] + \boldsymbol{\epsilon}.
\end{equation}
$\mathcal{M}_k$ returns only the subset of system variables corresponding to the observed floe positions and orientations at time $t_k$.
The dependence on $k$ allows for a changing number of observed floes at each observation time.
$\boldsymbol{\epsilon}$ is a small Gaussian observational noise, corresponding to the resolution of the satellite images.

Let $N$ be the size of the ensemble and denote individual ensemble members by $\boldsymbol{\psi}^{(n)}_k(t)$ for $n = 1, \dots, N$.
The superscript ``$(n)$'' distinguishes individual ensemble members from the true system state denoted by $\boldsymbol{\psi}(t)$.
The subscript $k$ denotes that the ensemble has been updated using the first $k$ observations.
To compute $\boldsymbol{\psi}^{(i)}_k(t)$ for $t > t_k$, the forecast model is used.
While only the value of $\boldsymbol{\psi}^{(i)}_k(t_{k+1})$ is required to perform the ensemble update, the ensemble at any prior time $t$ for $t_1 \leq t \leq t_{k+1}$ can be stored in memory and updated for each new observation.

Once the ensemble has been updated for observation $k$, to assimilate observation $k+1$, the $M \times N$ matrix of the ensemble members is formed
\begin{equation}
\boldsymbol{A}_k(t) = \left(\begin{matrix} \boldsymbol{\psi}^{(1)}_k(t) & \boldsymbol{\psi}^{(2)}_k(t) & \cdots & \boldsymbol{\psi}^{(N)}_k(t) \end{matrix} \right).
\end{equation}
The forecast ensemble matrix, $\boldsymbol{A}^\mathrm{f}_{k+1} = \boldsymbol{A}_k(t_{k+1})$, is calculated using the forecast model.
Then the updated ensemble is calculated using an $N \times N$ linear transformation, $\boldsymbol{T}$, of the ensemble
\begin{equation}
    \boldsymbol{A}_{k+1}(t) = \boldsymbol{A}_k(t) \boldsymbol{T}
\end{equation}
for any $t_1 \leq t \leq t_{k+1}$ where $T$ is formed using the Kalman filter equations from the forecast ensemble, $\boldsymbol{A}^\mathrm{f}_{k+1}$, and the observations, $\boldsymbol{d}_{k+1}$.

The above method is modified slightly to utilize localization, which leverages the spatial structure of the system to reduce the negative effects of spurious correlations.
During the update at time $t_k$, each variable in $\boldsymbol{\psi}$ is associated with a location in physical space.
For variables of an observed floe, their assigned location is the position of the observation.
For an unobserved floe, the forecast mean position is used.
For ocean and atmosphere variables, the spectral representations are transformed to physical space and the location for each grid point is used.

To update under localization, each state variable in $\boldsymbol{\psi}$ is updated individually.
An ensemble corresponding to each state variable is formed from taking the values from the ensemble of model trajectories.
This ensemble is then updated using only observations that are within a fixed radius of that variable's associated spatial location.
The matrix $\boldsymbol{T}$ is formed using the full forecast ensemble, $\boldsymbol{A}^\mathrm{f}_k$, and the observation vector $\boldsymbol{d}_k$ containing only the observations within the localization radius.
Then the $1 \times M$ row vector of the variable's ensemble is updated using this localized version of $\boldsymbol{T}$.
This process is repeated to update all localized variables in $\boldsymbol{\psi}$.

\subsection{Parameter Estimation}

Unknown parameters, such as individual floe thicknesses, can be estimated within the proposed framework.
The parameters are appended to the state vector and treated as non-dynamical variables. In other words, the evolution equations of the parameters $\boldsymbol\theta$ are $\d\boldsymbol\theta/\d t=0$.
The initial values of the parameters in each ensemble member are drawn from a background distribution (displayed in the right panel of Figure \ref{fig:thickness}). Therefore, the initial ensemble includes the uncertainty of the parameters.
During the ensemble forecast, the parameter values are kept constant.
However, during the ensemble update, the distribution of parameter values among the ensemble members is updated as well according to the same linear transformation.
In this way the distribution of parameter values changes over the course of the algorithm even though the values do not change during the forecast.
Since the parameters are non-dynamical, trajectories of these parameters do not need to be considered and only the current values of the parameters need to be stored.

\bmhead{Acknowledgments}

The research of N.C. was partially funded by Office of Naval Research (ONR) Multidisciplinary University Initiative (MURI) award N00014-19-1-2421.  M.M.W. was funded by the ONR awards N00014-20-1-2753 and N00014-19-1-2421. J.C. was supported as research assistant under this grant and by the National Science Foundation award DMS-2023239 through the Institute for Foundations of Data Science (IFDS) at UW-Madison. The authors gratefully acknowledge Dr. Georgy Manucharyan for insightful discussions, and Dr. Rosalinda Lopez-Acosta for her work on the development of the Ice Floe Tracker algorithm.

\backmatter

\bmhead{Supplementary information}

The submission contains supplementary material.
\bmhead{Data availability} The public data sets used in the findings of this study can be found in the references within the paper. The data needed to reproduce the results can be found in the below-mentioned GitHub repository.

\bmhead{Code availability} The codes are written in MATLAB and are available on GitHub (\url{https://github.com/JeffreyCovington/floe-interpolation})

\bmhead{Author contributions}
N.C. and M.M.W. designed the research. M.M.W. provided the satellite ice floe observations. J.C. and N.C. developed the methodology. All authors performed the research and contributed to the manuscript.

\bmhead{Competing interests}
The authors declare no competing interests.



\end{document}

\noindent{\Large \emph{Supporting Information:}}\medskip

\begin{center} 
{\Large Bridging Gaps in the Climate Observation
Network: \\A Physics-based Nonlinear
Dynamical Interpolation of Lagrangian Ice
Floe Measurements via Data-Driven
Stochastic Models}\medskip

Jeffrey Covington, Nan Chen and Monica M. Wilhelmus
\end{center}

\section{Sensitivity Analysis of the Ocean Recovery}

The ocean is represented in the system state by the stream function, which encodes an incompressible flow.
The synthetic data was generated using a quasi-geostrophic (QG) incompressible ocean model.
For nonlinear data assimilation, a statistically accurate stochastic forecast model is used to forecast the ensemble of model trajectories.
This ensemble is then updated after each set of observations, allowing any model variables, including the ocean stream function, to be recovered at any point in time.
For these experiments, the ocean is recovered for the day of July 1, which is roughly in the middle of the set of observations.
Note that the ocean evolves slowly with time and so this is representative of the overall ocean recovery. In addition to the standard setup in the main text, the following two additional experiments of the ocean flow field recovery are included.

\begin{figure}[h]
\centering
\includegraphics[width=9cm]{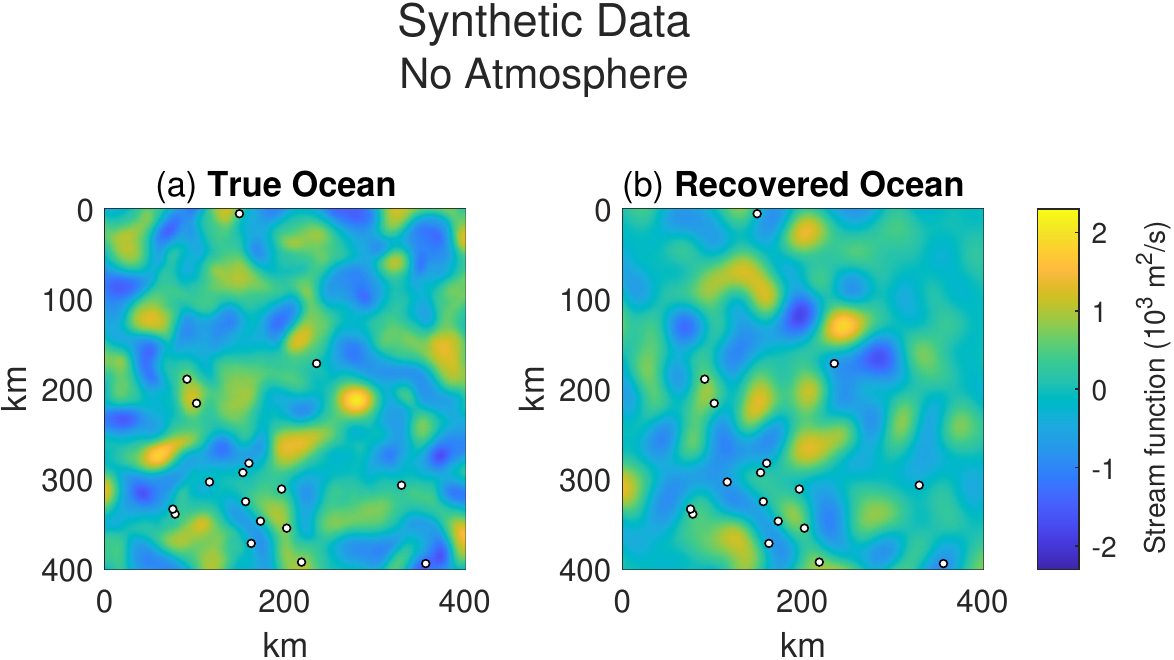}
\caption{A comparison of the true and recovered ocean on July 1 for a synthetic data experiment with no atmosphere forcing. Panel (a) shows the stream function of the true ocean generated from a QG ocean model. Panel (b) shows the recovery of the stream function using the ensemble mean. Each plot shows white dots indicating the position of the observed floes. While a 400 km by 400 km region is plotted, the ocean is extends another 100 km on all sides to reduce the impact of the periodicity in the ocean model.}\label{fig:no_atmosphere}
\end{figure}

\begin{figure}[h]
\centering
\includegraphics[width=9cm]{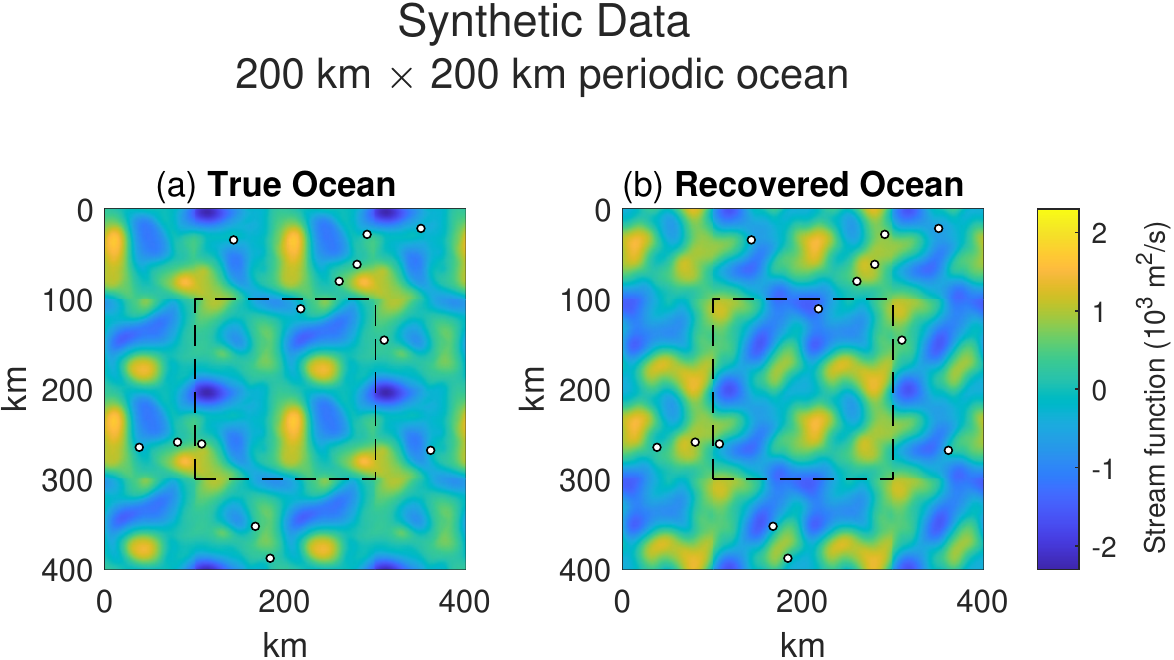}
\caption{Compares the true and recovered ocean stream functions for a synthetic data experiment using a 200 km by 200 km ocean extended periodically to the whole domain. Panel (a) shows the true ocean, generated from a QG model on a 200 km by 200 km domain. The dashed black line indicates the size of the ocean which is then periodically extended to the whole domain. Panel (b) shows the ensemble mean stream function. The floe positions are indicated with white dots. Note that observations outside of the 200 km by 200 km ocean region influence the ensemble update of the ocean just as much as observations inside the region due to the periodicity of the ocean.}\label{fig:200_domain}
\end{figure}

Figure \ref{fig:no_atmosphere} shows the recovery of the ocean stream function using synthetic data without atmosphere forcing, i.e. atmosphere component of the sea ice model is removed for both the generation of synthetic data and in the forecast model.
In this case the dynamics of the floes are fully accounted for by ocean forces and the total degree of freedom of the unobserved variables is reduced.
Therefore, a better recovery of the ocean state is expected due to the lack of interference from the atmosphere.
Indeed, the results of the ocean recovery seem to be improved.
However, the lack of atmosphere forcing, which is responsible for the large-scale floe movements, leads to floe trajectories that travel a shorter distance, and hence cover a smaller portion of the region.
This impacts localization during the ocean ensemble update, which removes the influence of observed floes that fall outside the localization radius of each ocean grid point.
In this case the amplitude of the recovered ocean is nearly zero in some areas, which reflects the uncertainty of the ensemble rather than an actual amplitude of individual ensemble members.

In another experiment, Figure \ref{fig:200_domain} shows the ocean recovery for synthetic data that is generated with atmospheric forcing, but using a 200 km by 200 km ocean (rather than a 600 km by 600 km ocean) that is extended periodically to the entire domain.
This is analogous to a case in which a much higher density of observations is available.
No localization is utilized here, as the typical localization radius of 200 km covers the entire ocean.
In this case the qualitative recovery of the ocean is quite good, demonstrating that a sufficiently high density of floe observations can recover ocean features well.

\begin{comment}
\section{Thickness Estimation}

Estimation of individual floe thicknesses is accomplished through parameter estimation.
The number of available observations for a floe trajectory is key to the skill of thickness estimation.
In the main experiments, the number of available observations was constrained by the limitations of the real data set.
Figure \ref{fig:thicknessfig_short} shows an example where synthetic data was generated with 20 days of observations for each floe.
The thickness estimation is much improved, both in the ensemble mean and also the reduction in the ensemble spread, representing higher confidence.

\begin{figure}[h]
\centering
\includegraphics[width=\textwidth]{Figs/thicknessfig_short.pdf}
\label{fig:thicknessfig_short}
\caption{Shows the estimation of floe thickness using synthetic data with long trajectories of 20 days for each of 20 floes.}
\end{figure}
\end{comment}

\section{Behavior of Ensemble Members in the Dynamical Interpolation}

%In this framework, observations of daily floe positions are dynamically interpolated using the ensemble mean.
The ensemble mean provides a complete trajectory and is the best point estimate of a floe's position available from the dynamical interpolation method. However, despite the fact that each ensemble member is based on the model forecast corrected by the partial observations, their average --- the ensemble mean --- is often not a physically consistent trajectory of the model. Therefore, it is important to look at each individual ensemble member and understand the additional physical properties that are not fully reflected in the ensemble mean.

%From the ensemble update, each ensemble member satisfies the observed floe locations, within a small observational error. Thus, individual sampled ensemble members provide an alternative dynamical interpolation scheme to the ensemble mean. While still not true trajectories of this nonlinear model, the sampled ensemble trajectories may better represent physical properties of true floe trajectories.

\begin{figure}[h]
\centering
\includegraphics[width=7cm]{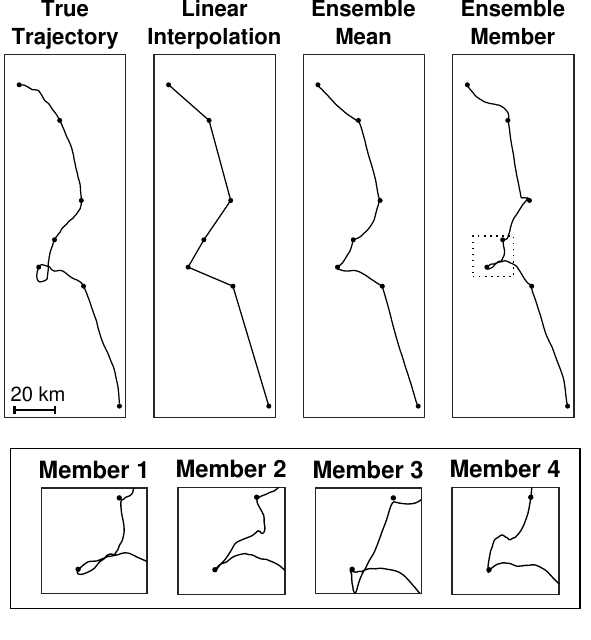}
\caption{Comparison of different methods for the interpolation of floe trajectories. The first panel shows a floe trajectory generated from the sea ice model and sampled every time unit at the black dots. This trajectory contains a loop. The next three panels show various forms of interpolation. The first shows linear interpolation. The second shows the ensemble mean. The third shows a sampled ensemble member. The bottom panels shows other sampled ensemble members.}\label{fig:trajectory_comparison}
\end{figure}

Figure \ref{fig:trajectory_comparison} shows an example of a continuous floe trajectory generated from the sea ice model, along with the daily observed floe positions.
The trajectories between daily observations are then interpolated using different methods.
Note that this floe trajectory contains a loop midway through, which is compared qualitatively with the interpolation using different methods.
While the ensemble mean provides better physics than linear interpolation and contains nonlinear evolution of the trajectory, it still fails to capture this loop.
On the other hand, more than half of the ensemble members recover such a loop around the correct location, something which is otherwise lost during averaging.
The finding here reveals the potential role that utilizing individual ensemble members can play in identifying properties of floe trajectories that are not otherwise captured by the ensemble mean. Such a result also highlights the importance of considering the uncertainty represented by the ensemble members in addition to the mean state estimation.

